\documentclass[twocolumn,english,prX,,amsmath,longbibliography]{revtex4-1}

\usepackage[T1]{fontenc}
\usepackage{graphicx}
\usepackage{dcolumn}
\usepackage{bm}
\usepackage{subfig}
\usepackage{array}
\usepackage{pdfpages}

\newcommand{\beq}{\begin{equation}}
\newcommand{\eeq}{\end{equation}}
\newcommand{\beqar}{\begin{eqnarray}}
\newcommand{\eeqar}{\end{eqnarray}}
\newcommand{\bea}{\begin{eqnarray}}
\newcommand{\eea}{\end{eqnarray}}
\newcommand{\bcen}{\begin{center}}
\newcommand{\ecen}{\end{center}}

\newcommand{\ket}[1]{\left| #1 \right>}

\newcommand{\tr}{\mathrm{tr}}
\newcommand{\ave}[1]{\left< #1 \right>}
\newcommand{\dg}{^\dagger}
\newcommand{\Hc}{\mathrm{H.c.}}
\newcommand{\Herm}{\mathcal{H}}

\newcommand{\Mean}{\mathbf{M}}

\newcommand{\half}{\frac{1}{2}}
\newcommand{\Hop}{\hat H}
\newcommand{\Kop}{\hat K}
\newcommand{\rop}{\hat\rho}
\newcommand{\ropo}{\hat\rho}
\newcommand{\sop}{\hat{\sigma}}
\newcommand{\aop}{\hat{a}}
\newcommand{\bop}{\hat{b}}
\newcommand{\cop}{\hat{c}}
\newcommand{\omh}{\omega_h}
\newcommand{\omc}{\omega_c}
\newcommand{\omo}{\omega_o}
\newcommand{\Lch}{\mathcal{L}_h}
\newcommand{\Lcc}{\mathcal{L}_c}
\newcommand{\Lce}{\mathcal{L}_e}
\newcommand{\Lcm}{\mathcal{L}_m}
\newcommand{\Lcf}{\mathcal{L}_f}
\newcommand{\betah}{\beta_h}
\newcommand{\betc}{\beta_c}
\newcommand{\beteneg}{\beta_e^-}

\newcommand{\epsd}{\epsilon_d}
\newcommand{\kapeneg}{\kappa_e^-}
\newcommand{\kapf}{\kappa_f}
\newcommand{\gamm}{\gamma_m}

\newcommand{\cbar}{\bar{c}}
\newcommand{\gsqrt}{\sqrt{\gamma_m}}
\newcommand{\Ener}{\mathcal{E}}
\newcommand{\Work}{\mathcal{W}}
\newcommand{\Heat}{\mathcal{J}}
\newcommand{\Pow}{\mathcal{P}}

\begin{document}

\title{Quantum Flywheel}

\author{Amikam Levy$^{(1)}$, Lajos Di\'{o}si$^{(2)}$ and Ronnie Kosloff$^{(1)}$ }

\affiliation{$^{(1)}$Fritz Haber Research Center for Molecular Dynamics, The Institute  of Chemistry,  The Hebrew University, Jerusalem 91904, Israel}
\affiliation{$^{(2)}$ Wigner Research Center for Physics, H-1525 Budapest 114. P.O. Box 49, Hungary }

\begin{abstract}

A quantum flywheel is studied with the purpose of storing useful work in quantum levels, while additional power is extracted continuously from the device. 
The flywheel gains its energy form a quantum heat engine.
Generally, when a work repository is quantized the work exchange with the engine is accompanied with heat exchange, which may degrade the charging efficiency.
In the particular realization of a quantum harmonic oscillator work repository, quantum and thermal fluctuations dominates the dynamics. 
Quantum monitoring and feedback control are applied to the flywheel, as it is shown to be an essential part of stabilizing and regulating its state of operation, and bringing the system to a steady state. 
A particular balance between information gained by measuring the system and the information fed back to the system is found to maximize the charging efficiency.  
The dynamics of the flywheel are described by a stochastic master equation that accounts for the engine, the external driving, the measurement, and the feedback operations.

\end{abstract}  
\date{\today}

\maketitle
\section{INTRODUCTION}
A \emph{flywheel} is a device that stores kinetic energy in the rotational motion of the wheel and supplies it on demand. 
In many devices the flywheel is an essential component for extracting work from an engine.
The main tasks of a flywheel are twofold: transducing discrete energy into continuous power and storing useful \emph{work}.  
This energy reserve can be rapidly drained on demand, ultimately gaining more power than the power supply charging it can deliver.       
Miniaturizing heat engines and refrigerators received much attention in the past decade.
Experimental set-ups of such devices were constructed in the micrometers domain \cite{clemens12,steeneken11}, and recently the operation of a single-atom heat engine was reported \cite{rossnagel15}.
Many theoretical studies of these devices were extended to the quantum domain, concentrated on the study of efficiency, power extraction and the study of thermodynamic laws (see reviews  \cite{levy14a,goold15,janet15,millen16,gelbwaser15b} and references therein).  
Work extraction from quantum systems was also studied extensively \citep{alahverdyan04,alicki13,skrzypczyk14}. 
\par
Any realistic engine is regulated by monitoring and a feedback loop. 
The purpose is to control its timing, adjust its frequency and amplitude to match the other parts of the device, and  compensate for unpredictable disturbances.
Recent theoretical studies demonstrated that quantum properties such as coherence and correlations enhance the work  extracted from the system \cite{uzdin15,marti15,korzekwa15,scully03}. 
Future quantum technologies aiming to exploit these quantum features will encounter the issue of regulating the device. 
Standard ideal quantum measurements will demolish these features.
Therefore, to overcome this problem a conceivable approach to regulate the quantum device is by continuous weak measurements (\emph{monitoring}) and \emph{feedback control}.
Another fundamental problem which is demonstrated in this study and that is resolved by monitoring and feedback control is the unlimited entropy increase of the work repository, i.e. proliferating fluctuations catastrophically heat up the flywheel. 
\par
In this paper we introduce the concept of a quantum flywheel as an integral part of a quantum \emph{heat engine}. 
The flywheel is composed of a quantum harmonic oscillator (HO) interacting with a two qubit quantum heat engine.
If the flywheel was simply replaced by an external classical field then the engine would operate in steady state and power can continuously be extracted from the engine(see Appendix E for details). 
However, if the task is to \emph{store} work in a quantum HO by coupling it to the engine,  the HO will be subject to a fatal growth of fluctuations and establishment of \emph{steady state} is impossible.  
The HO is unstable even when an external \emph{driving} field is utilized to extract power and stabilize it. 
By applying monitoring and feedback control we obtain a steady state operation for the flywheel, continuously gaining power, and storing useful work in the flywheel that later can be extracted.
\par
Monitored and controlled quantum heat engines are still to be realized experimentally, however, the individual components already exist.
Quantum monitoring and feedback control experiments exists for various HO's such as electromagnetic cavity, nanomechanical oscillators, trapped particles and superconducting circuits, see review \cite{nori14} and reference therein.
Single microscopic quantum heat engines realisations are still under development with only few examples as of today \cite{koski14,rossnagel15}. 
\section{Heat Engine Operation}
\begin{figure}
\centering
\begin{tabular}{cc}
\subfloat[]{\includegraphics[height=3cm]{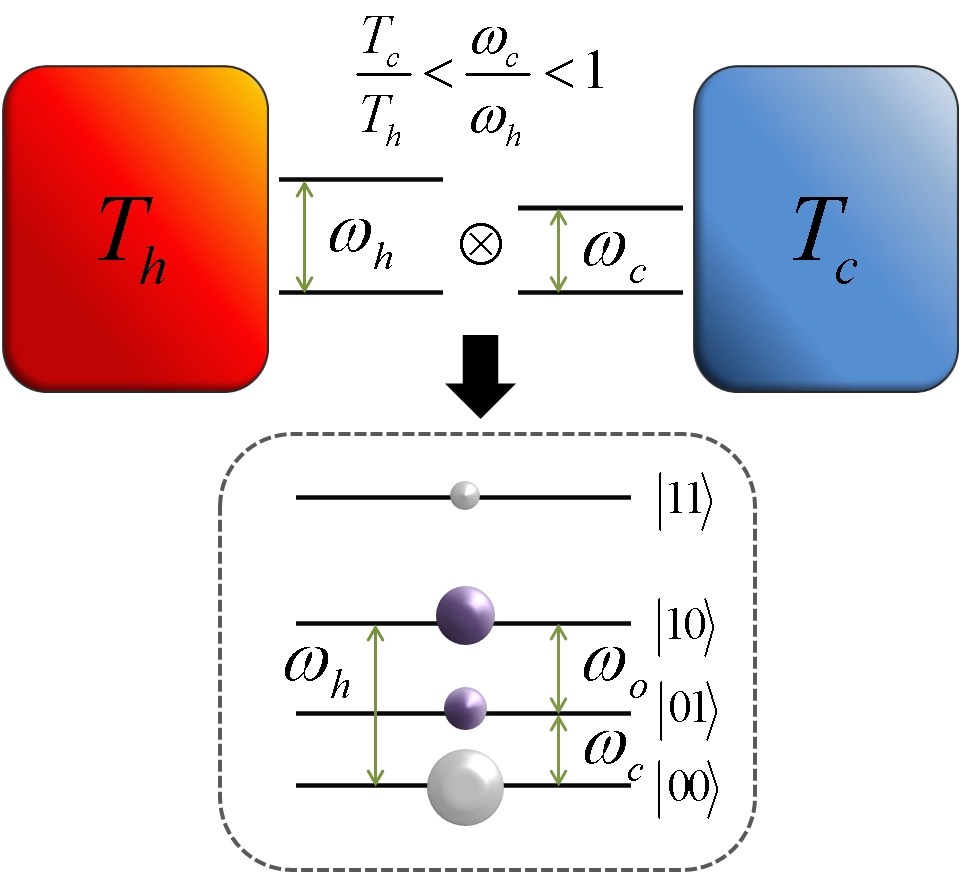}}&
\subfloat[]{\includegraphics[height=3cm]{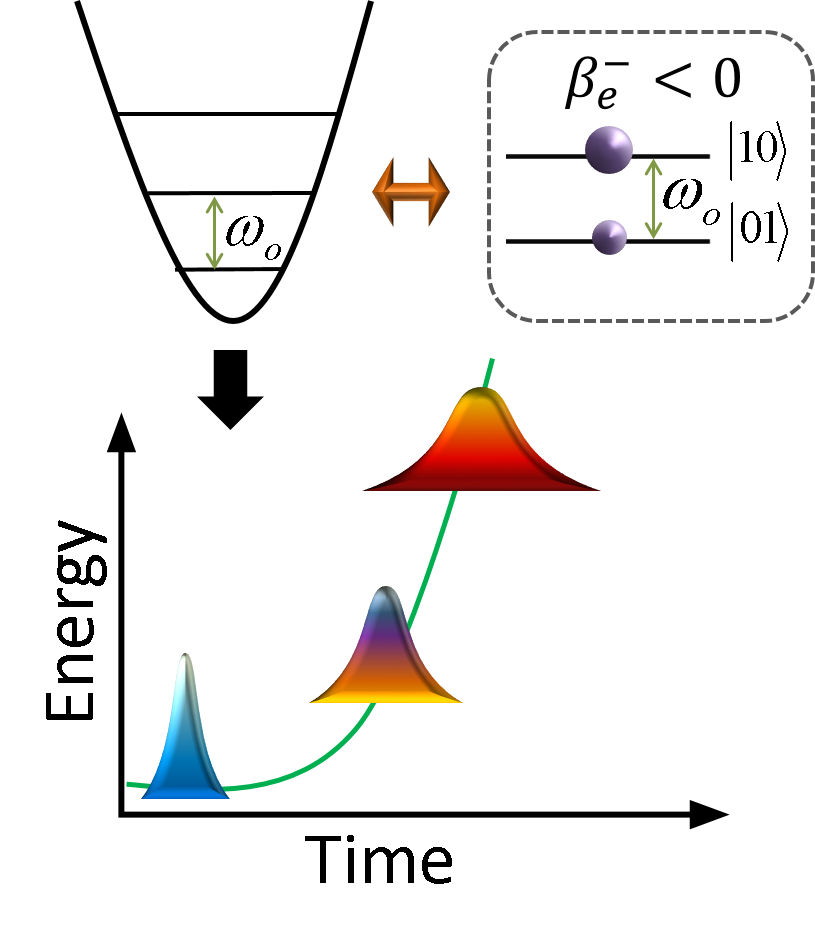}}
\\
\subfloat[]{\includegraphics[height=3cm]{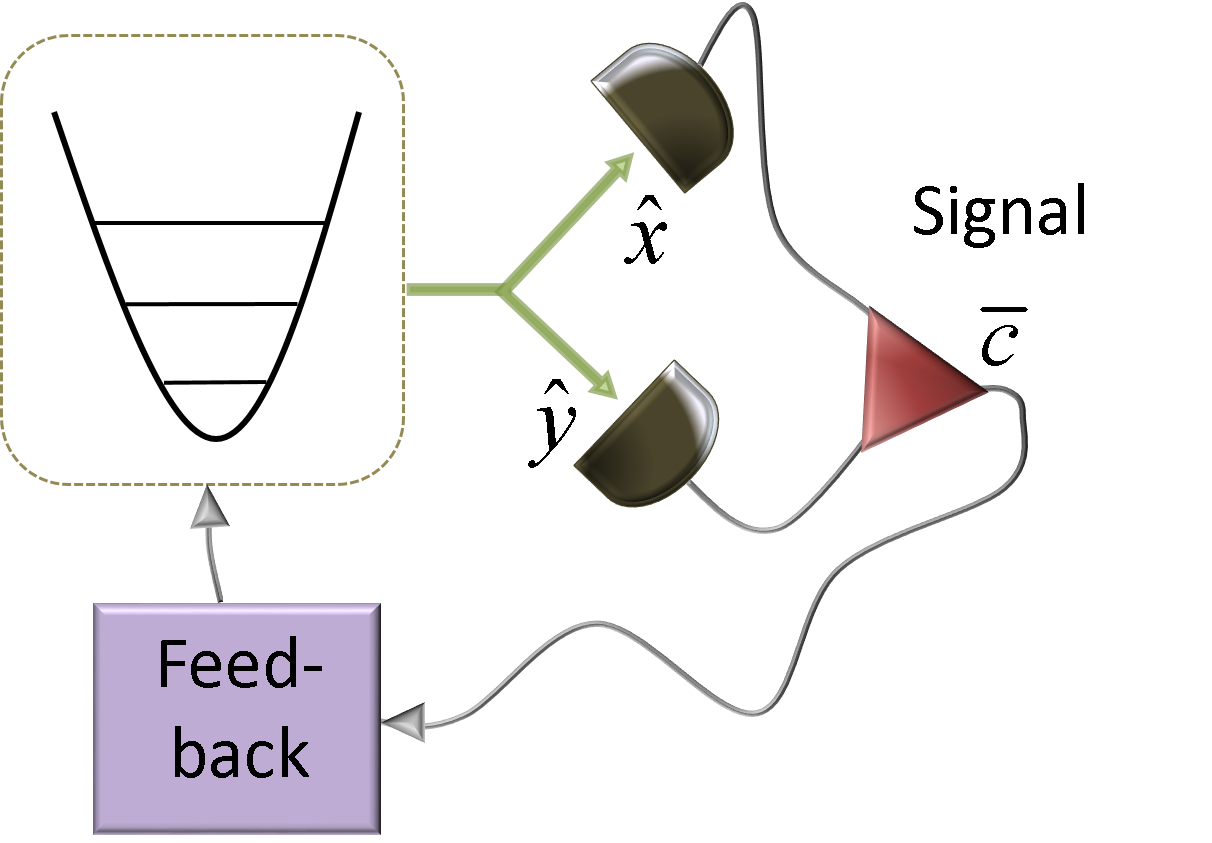}}&
\subfloat[]{\includegraphics[height=3cm]{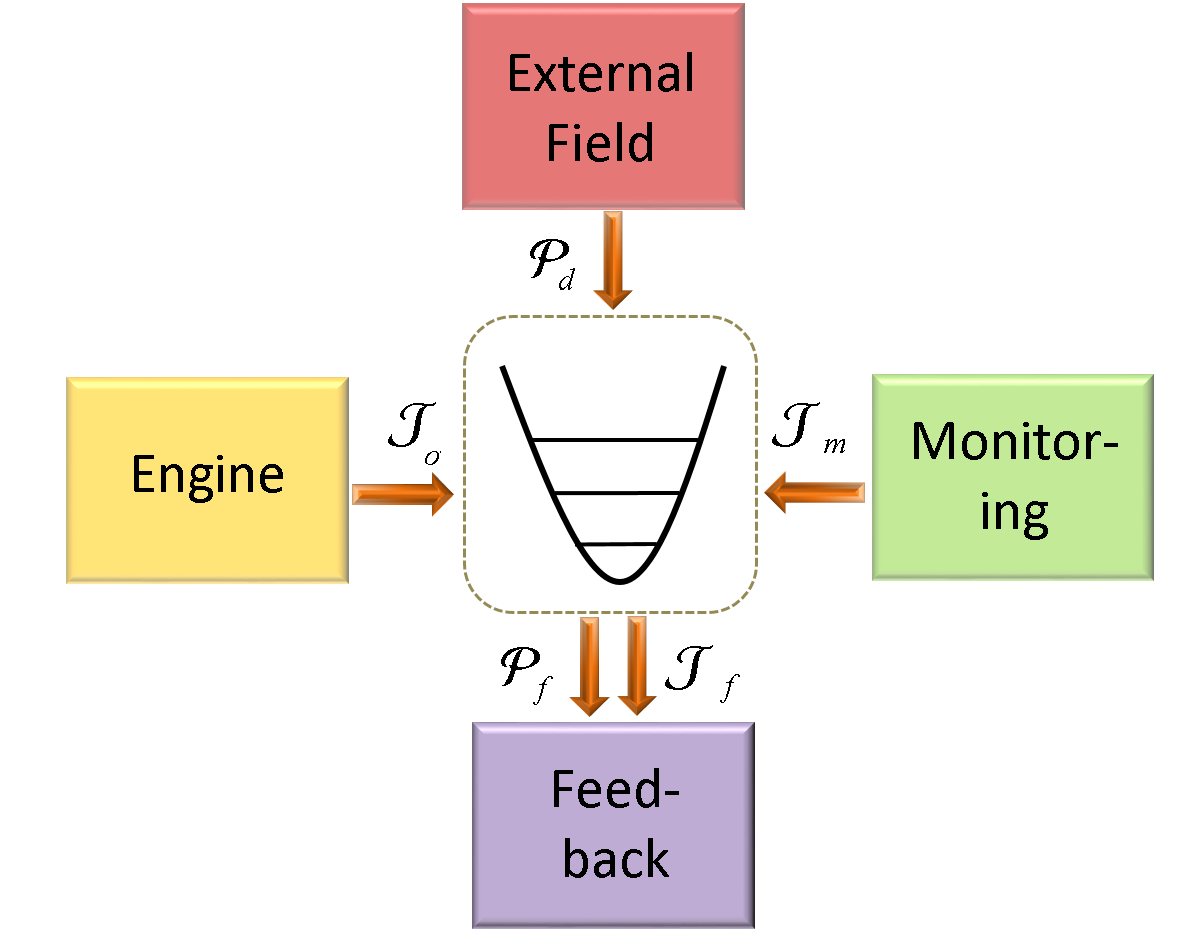}}
\end{tabular}
\caption{Scheme of the different components and operations of the heat engine with a flywheel.
(a) The state of two qubits of the heat engine, coupled to heat baths at temperatures $T_h$ and $T_c$, is represented as a two-qubit state with population inversion between the 2nd and 3rd energy levels. The size of the sphere represents the population in each level. 
(b) The population inversion in the engine corresponds to a heat bath with the inverse negative temperature $\beteneg$. This bath is coupled to the harmonic oscillator (flywheel), increasing exponentially its energy and the width of phase-space probability distribution. 
(c) Measurement of the quadratures of the harmonic oscillator, resulting in the signal $\bar{c}$. The signal is then fed back to the oscillator to ensure a steady-state. 
(d) Energy flow chart of the different components in steady-state of the flywheel.}
\label{fig:schem}
\end{figure}

 The basic concept of a quantum heat engine (similar to the classical one) is consists of two thermal heat baths at different temperature, a working medium, and a work repository.  
In the quantum counterpart the working medium is quantized and the work repository can be either an external classical field \cite{levy14a} or it can be quantized as well \cite{gelbwaser15}. 
Here we consider the operation of a continuous quantum engine for which the heat baths and the work repository are coupled simultaneously and continuously to the working medium \cite{levy14a}. 
The working medium is comprised of two qubits, with the Hamiltonians $\hat{H}_a=\omega_h\aop^{\dagger}\aop$ and $\hat{H}_b=\omega_c\bop^{\dagger}\bop$ respectively. 
Each qubit is weakly coupled to a different heat bath with the inverse temperature $\beta_h$ and $\beta_c$, where the indexes $h$ and $c$ stand for $hot$ and $cold$. 
The dynamics of the qubits follow the standart thermalizing master equation of Lindblad-Gorini- Kossakowski-Sudarshan (LGKS) \cite{lindblad76,gorini276}. 
The asymptotic two-qubit state $\rop_h^\infty\otimes\rop_c^\infty$ is the product of the thermal equilibrium Gibbs states of the two qubits, respectively at hot and cold temperatures $1/\betah$ and $1/\betc$. 
Satisfying the heat engine conditions, $\beta_h/\beta_c < \omega_c/\omega_h<1$,  population inversion is obtained between the 3rd level $\ket{10}$ and the 2nd level $\ket{01}$ (see Fig.\ref{fig:schem}a). 
The populations of these states are given by $p_{10}=n_h(1-n_c)$ and  $p_{01}=n_c(1-n_h)$. 
Here, $n_{h(c)}=\left[\exp(\beta_{h(c)}\omega_{h(c)})+1\right]^{-1} $  are the thermal occupation numbers in $\rop_{h(c)}^\infty$.
The 2nd and the 3rd levels are treated as an effective two level system (TLS) with the energy gap $\omega_o = \omega_h-\omega_c $ (we take $\hbar=k_B=1$). 
The state of this TLS is a Gibbs state with a negative effective temperature
\begin{equation}
\frac{1}{\beteneg}=\frac{\omega_h-\omega_c}{\beta_h\omega_h-\beta_c\omega_c}<0.
\end{equation}
\par
We exploit the TLS population inversion to "charge" a quantum harmonic oscillator (HO) with useful work.
The Hamiltonian of the HO and the TLS-HO interaction Hamiltonian are given by $\Hop_o=\omega_o \cop^{\dagger}\cop$, and $\hat{K}=ig(\aop^{\dagger}\bop\cop- \aop\bop^{\dagger}\cop^{\dagger})$, respectively.
Given that the termalization time of the qubits is much shorter then the internal time scale, then
\beq
\label{eq:weak_coupling}
g\sqrt{\ave{\cop^{\dagger}\cop}+1}\ll\Gamma_{h(c)} \left(1+\exp(-\beta_{h(c)}\omega_{h(c)})\right),
\eeq
the TLS can be considered heuristically as a heat bath with negative temperature weakly coupled to the HO. We prove that indeed the state $\rop$ of the HO  satisfies the standard thermalizing master equation extended to negative temperature $1/\beteneg$, which in the interaction picture of $\Hop_o$ takes the form
\begin{equation}
\label{eq:ME_e}
\frac{d\rop}{dt}=\Lce\rop\equiv
\Gamma_e(\cop\rop\cop\dg-\Herm\cop\dg\cop\rop)
      +\Gamma_e e^{-\beteneg\omo}(\cop\dg\rop\cop-\Herm\cop\cop\dg\rop).
\end{equation}
The damping rate $\Gamma_e$ is proportional to the squared coupling $g^2$, and depends on the parameters of the engine, such as the occupations $n_{h(c)}$ and the rates $\Gamma_{h(c)}$ (see Appendix A). 
The notation $\mathcal{H}$ stands for the Hermitian part of everything coming after it (different from the convention in  ref \cite{wisemanbook}). 
A rigorous derivation of equation (\ref{eq:ME_e}) can be found in Appendix A. 
\par
Since $\beteneg < 0$ the master equation \eqref{eq:ME_e} has no steady state solution, energy will constantly flow into the flywheel.
The parameters containing the superscript $^-$ are negative.
The standard equations remain valid for the
mean amplitude $\ave{\cop}_t$ and the occupation $\ave{\cop^{\dagger}\cop}_t$:
\begin{eqnarray}
\label{eq:c_t}
\frac{d\ave{\cop}_t}{dt}&=&-\left(\kapeneg+i\omo\right)\ave{\cop}_t\\
\label{eq:n_t}
\frac{d\ave{\cop^\dagger\cop}_t}{dt}&=&-2\kapeneg\ave{\cop^\dagger\cop}_t+\Gamma_e e^{-\beteneg\omo}
\end{eqnarray} 
where the amplitude damping rate 
\beq
\label{eq:kappa_neg}
\kapeneg = \half\Gamma_e\left(1-e^{-\beteneg\omo}\right)
\eeq
takes negative values since $\beteneg<0$ (see Appendix A). 
Therefore both $\ave{\cop}_t$ and $\ave{\cop^\dagger\cop}_t$ (and all higher moments) diverge exponentially with time (see Fig.\ref{fig:schem}b)
resulting in the instability of the dynamics against small perturbations.
In particular, an initial Gibbs state maintains its form but with an exponentially growing temperature 
$1/\beta_t=\omega_o/\log\left(1+ \ave{\cop^{\dagger}\cop}_t^{-1} \right)$. 
Thus, $\rop(t)\propto\exp\left(-\beta_t\omega_o\cop^\dagger\cop\right)$
is an unstable solution of the master eq. \eqref{eq:ME_e}. Any small
perturbation will divert it from the class of Gibbs states. A more general
class of solutions, displaced Gibbs states    
\begin{equation}
\rop(t)\propto\exp\left[-\beta_t\omega_o
(\cop-\ave{\cop}_t)^\dagger(\cop-\ave{\cop}_t) \right],\nonumber
\end{equation}
with effective temperature
 $1/\beta_t=\omega_o/\log\left[1+ \left( \ave{\cop^{\dagger}\cop}_t-|\ave{\cop}_t|^2\right)^{-1} \right]$ will, in principle, be suitable for work extraction.
But this option is misleading since the instability of the above solutions is not yet resolved.  
\par
A reasonable approach to stabilize the flywheel while extracting additional power is achieved by driving the HO via a resonant oscillating external field. 
The field is expressed by the time dependent Hamiltonian, $\Hop_d(t)=-i\epsd\cop^{\dagger}e^{-i\omega_o t}+\Hc$.
The master equation in the interaction picture \eqref{eq:ME_e} becomes modified by a static Hamiltonian term (see Appendix B):
\begin{equation}
\label{eq:ME_e_d}
\frac{d\rop}{dt}=\Lce\rop-\epsd\left[\cop^{\dagger} -\cop,\rop_o\right].
\end{equation}
Indeed, equation (\ref{eq:ME_e_d}) leads to a stationary amplitude with a rotating phase:
$\ave{\cop}_t=-(\epsd/\kapeneg)e^{-i\omega_o t}$.
Nevertheless, the stationary state remains unstable,
the occupation number and higher moments  diverge invariably.    
Driving in itself can not solve the instability issue.
Unlimited growth of quantum and thermal fluctuations must be
suppressed by active control of the flywheel.

\section{Measurement and Feedback Control}
Continuous measurement, i.e., monitoring, is  the first task towards implementing feedback control \cite{wisemanbook}. 
By applying monitoring and feedback control we can stabilize the  flywheel and charge it with useful work. 
Consider a time-continuous measurement of  both quadratures $\hat{x}=\frac{1}{\sqrt{2}}(\cop^\dagger +\cop)$ and $\hat{y}=\frac{i}{\sqrt{2}}(\cop^\dagger -\cop)$ of the HO.
Generalizing the result of \cite{diosi88}, we simultaneously monitor $\hat{x}$ and $\hat{y}$ (see Fig.\ref{fig:schem}c). 
The dynamics is described by a stochastic master equation (SME) for the density operator $\sop$ conditioned on both measurement signals $\bar{x},\bar{y}$ (see Appendix C). 
The stochastic mean ($\Mean$) of the conditional state yields the unconditional state, i.e., $\Mean\sop=\rop$ satisfying a corresponding master equation of the usual LGKS structure. 
It differs from the master equation eq.(\ref{eq:ME_e_d}) 
by the additional monitoring term 
\begin{equation}
\label{eq:L_m}
\Lcm\rop= \frac{\gamm}{4}(\cop\rop\cop\dg-\Herm\cop\dg\cop\rop + \cop\dg\rop\cop-\Herm\cop\cop\dg\rop),
\end{equation} 
 where $\gamm$ is the measurement strength. This generator corresponds to an infinite temperature heat bath. 
Hence, the act of monitoring additionally heats the flywheel, and
contributes to the undesirable proliferating fluctuations of the HO.
\par
Stabilization is accomplished by a feedback loop conditioned on the measured signals $\bar{x},\bar{y}$. 
As a result, the HO is kept in the vicinity of the constant rotating amplitude set by the external driving. 
The feedback Hamiltonian in the  Schr\"{o}dinger picture is given by $\Hop_f(t)=-i\kapf\cbar(t)\cop\dg +\Hc$, where $\bar{c}=\frac{1}{\sqrt{2}}(\bar{x}+i\bar{y})$ is the complex representation of the two real signals $\bar{x}$ and $\bar{y}$, and $\kapf$ is the feedback strength. 
By setting the value of $\kapf$ the steady-state of the flywheel is guaranteed. 
The feedback is applied on top of the monitored evolution, 
$\sop+d\sop\rightarrow e^{-i\Hop_f dt}(\sop+d\sop) e^{i\Hop_f dt}$,
yielding a SME for the conditional state (see Appendix D for details).
The master equation of the unconditional state reads
\begin{equation}
\label{eq:ME_e_d_m_f}
\frac{d\rop}{dt}=(\Lce+\Lcm+\Lcf)\rop-\epsd\left[\hat{c}^{\dagger}-\hat{c},\rop\right].
\end{equation}
The dissipative contribution of the feedback is:
\begin{eqnarray}
\label{eq:L_f}
\Lcf\sop=
 \left(\frac{\kapf^2}{\gamm}+\kapf\right)\left(\cop\sop\cop\dg-\Herm\cop\dg\cop\sop\right)
 \\  \nonumber
 + \left(\frac{\kapf^2}{\gamm}-\kapf\right)\left(\cop\dg\sop\cop-\Herm\cop\cop\dg\sop\right).
\end{eqnarray}

For $\kapf>\gamm$ this corresponds to a thermal bath of positive temperature.  Entering the regime $0<\kapf<\gamm$,
the cooling effect of $\Lcf$ within the sum  $\Lce+\Lcm+\Lcf$ becomes enhanced although 
$\Lcf$ ceases to be a mathematically correct dissipator in itself.
Eq.(\ref{eq:ME_e_d_m_f}) can be written in a compact form,
\begin{eqnarray}
\label{eq:ME_final}
\begin{array}{lr}
\frac{d\rop}{dt}=
 \Gamma\left(\cop\rop\cop\dg-\Herm\cop\dg\cop\rop\right)
+\Gamma e^{-\beta\omo}\left(\cop\dg\rop\cop-\Herm\cop\cop\dg\rop\right)
\\ 
 ~~~ - ~\epsd[\cop\dg-\cop,\rop]
\end{array}
\end{eqnarray}
where $\Gamma$ and $\beta$ are determined by
\beq
\label{eq:rate_damp}
\Gamma=\Gamma_e+\frac{\gamm}{4}+\frac{\kapf^2}{\gamm}+\kapf
\eeq
\beq
\label{eq:rate_pump}
e^{-\beta\omo}\Gamma=\Gamma_e e^{-\beteneg\omo}+\frac{\gamm}{4}+\frac{\kapf^2}{\gamm}-\kapf.
\eeq
\par 
The effective temperature $1/\beta$ becomes positive by setting the feedback strength above the following threshold:  
\beq
\label{eq:kappa_thresh}
\kapf > -\kapeneg=\half\Gamma_e\left(e^{-\beteneg\omega_o}-1\right).
\eeq
To summarize, as a result of the feedback the negative temperature
heat bath and the negative amplitude damping rate $\kapeneg$ for HO, becomes an effective positive temperature heat bath with amplitude damping rate $\kapf+\kapeneg>0$.
\section{Steady State and Work Extraction}
For sufficient strong feedback $\kapf$, satisfying $\kapf+\kapeneg>0$, the master equation \eqref{eq:ME_final} is a standard thermalizing master equation with resonant external driving. 
It has a unique stationary state which in the Schr\"{o}dinger picture is a thermal state with rotating displacement (see Appendix D),
\beq
\label{eq:steady-Sch}
\rop^{\infty}\propto\exp\left[-\beta\omo\left(\cop-c_\infty e^{-i\omo t}\right)^{\dagger}\left(\cop-c_\infty e^{-i\omo t}\right)\right].
\eeq 
where  
\beq
\label{eq:c_infty}
c_\infty=-\frac{\epsd}{\kapf+\kapeneg}<0.
\eeq 
Hence, the mean amplitude rotates, $\ave{\cop}_\infty=c_\infty e^{-i\omo t}$, its phase is shifted by $-\pi/2$ with respect to the external driving .   
The average population is given by the sum of the Bose statistic  $n_o$ and the yield of displacement 
\beq
\label{eq:n_infty}
\ave{\cop^{\dagger}\cop}_\infty =  \frac{1}{e^{\beta\omo}-1} +\vert c_\infty\vert^2 \equiv n_o+\vert c_\infty\vert^2.
\eeq
\par
We distinguish two opposing regimes of the steady-state operation of the flywheel. 
The first is the deep quantum regime, $n_o,\vert c_\infty\vert^2\ll1$, where the flywheel is operating in the vicinity of its ground state. 
The second is the classical regime in which both the thermal occupation and the displacement are large  numbers, $n_o,\vert c_\infty\vert^2\gg1$.
The two crossed regimes also present peculiar quantum features.
Recall that weak coupling condition \eqref{eq:weak_coupling} sets an
asymptotic upper limit on the total occupation (\ref{eq:n_infty}). 
This implies asymptotic upper limits on the temperature $1/\beta$, excluding too high thermal occupations $n_o$, as well as on the driving strength $\epsd$, confining the displacements $\ave{\cop}_\infty$. 
Thus, accessibility to the classical regime depends on the physical properties of the two-qubit heat engine and its coupling $g$ to the flywheel. 
\begin{figure}[htbp]
\center{\includegraphics[width=8.6cm]{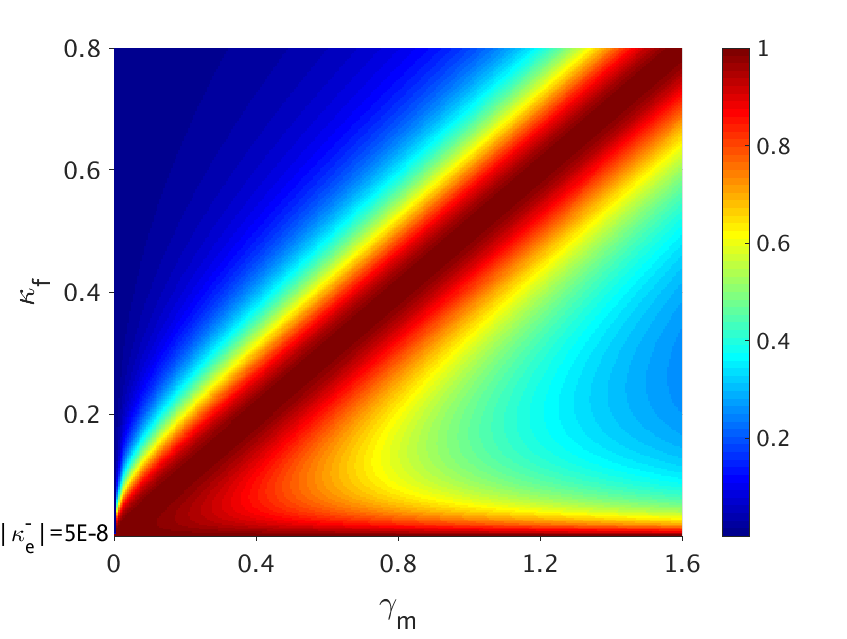}}
\caption{ Charging efficiency as function of measurement strength $\gamm$ and feedback strength $\kapf$.   The percentage of useful work out of the entire energy stored in the flywheel has a maximum for the ratio $\gamm / \kapf = 2$, and it is further maximised for $\kapf$ approaching its threshold $\vert\kapeneg\vert$.
Here: $\omo=1$, $\beteneg=-10^{-1}$, $\Gamma_e=10^{-6}$ and $\epsd=9\cdot10^{-2}$.}
\label{fig:Charging efficiency}
\end{figure}  
The steady  state \eqref{eq:steady-Sch} becomes a displaced Gibbs state and as such, it is suitable for work extraction.
The internal energy of the steady-state is given by 
$\Ener =  \omega_o(n_o+\vert c_\infty\vert^2)$. 
Applying a unitary displacement transformation can bring the state eq.(\ref{eq:steady-Sch}) into a Gibbs state (passive state) with the temperature $1/\beta$. Thus, the part of the internal energy that is due to $c_\infty$ can all be extracted by the unitary operation as the maximum useful work
\beq
\label{eq:Work}
\Work=\omega_o\vert c_\infty\vert^2=\frac{\omo\epsd^2}{(\kapf+\kapeneg)^2},
\eeq
which is independent of the strength $\gamm$ of the monitoring. 
The \emph{charging efficiency} of the flywheel can be defined as the ratio between useful work and the internal energy stored in the HO (see Fig.\ref{fig:Charging efficiency}),
\begin{equation}
\label{eq:Wefficiency}
\eta=\frac{\Work}{\Ener}=\frac{1}{1+n_o/\vert c_\infty\vert^2}.
\end{equation}
The efficiency is improved for small thermal occupation $n_o$ and large displacement $c_\infty$. 
The occupation $n_o$ becomes small when the effective temperature $1/\beta$ is reduced.
Interestingly, this singles out the optimum measurement strength $\gamm$ which has so far remained unconstrained. 
From eqs. (\ref{eq:rate_damp},\ref{eq:rate_pump}) we find that $1/\beta$ takes its minimum value with the choice $\gamm=2\kapf$ obtaining the expression 
\beq
n_o\vert_{min}=\frac{1}{(1+2\kapf/\Gamma_e)e^{\beteneg\omo}-1}
\eeq
of the minimum thermal occupation. The charging efficiency yields:
\beq
\eta\vert_{\gamm=2\kapf} 
=\frac{1}{1+\frac{\Gamma_e}{2\epsd^2}e^{-\beteneg\omo}(\kapf+\kapeneg)^2}.
\eeq
The efficiency $\eta_{\gamm=2\kapf}$ together with the extractable work $\Work$ reach higher values if we increase the displacement $\vert c_\infty\vert$.
In particular, the efficiency approaches its maximal value $1$
when the feedback $\kapf$ approaches its lower threshold, $\kapf \rightarrow -\kapeneg$. 
A different technique to maximize both the efficiency and the work is by increasing $\epsilon_d$, i.e., applying a stronger driving field.
Nevertheless, as was already mentioned, these two approaches are limited by the weak-coupling condition (\ref{eq:weak_coupling}).
\section{Energy Flows in Steady-State}
A macroscopic flywheel at rest requires an input work (initial push) to reach the vicinity of steady-state, at which point the output power is larger than the input power. 
Regulating the flywheel also has energetic costs that should be accounted for.
These energetic considerations, in principle, also apply to the quantum flywheel.
However, the related calculations would need a novel approach to  heat flow and power in quantum systems under stochastic control.
\par    
The standard definition of thermodynamic heat flow $\Heat$ and power 
$\Pow$ in open quantum systems is given \cite{alicki79} by the time derivative of the internal energy $\Ener=\tr[\rop \Hop]$ in the following manner:
\beq
\label{eq:dE}
d\Ener=\tr[d\rop \Hop]+\tr[\rop d\Hop]\equiv\Heat dt+\Pow dt.
\eeq
The Hamiltonian and the state of the system are typically stochastic in the theory of monitoring and feedback control.
Since stochastic fluctuations are microscopic, the thermodynamic definition of the internal energy is given by the stochastic mean  of the microscopic energy, $\Mean\tr(\sop\Hop)$.
This leads to the following generalization of the standard thermodynamic relation:
 \beq
 \label{eq:dE_stoch}
d\Ener=\Mean\tr[d\sop \Hop]+\Mean\tr[\sop d\Hop]\equiv\Heat dt+\Pow dt.
\eeq
The differentials in (\ref{eq:dE_stoch}) must be Stratonovich ones instead of those of Ito.
For the Ito differentials the rhs should contain the so-called Ito correction  $\Mean\tr[d\sop d\Hop]$ which would jeopardize the split of $d\Ener$ between heat flow and power.
\par
Any systematic calculation of heat flow and power requires to transform the final SME from Ito into Stratonovich form.
We postpone this very novel task to future research. 
Rather, we focus on the minimal calculations and considerations confirming that our model represents a genuine heat engine. 
\par
Next, we show that there is a consumable output power in the steady state operation of the flywheel.
The total Hamiltonian has two time-dependent contributions $H_d(t)$ and $H_f(t)$. 
Accordingly, the power $\Pow$ consists of two contributions corresponding to the power  invested by the driving  and the power  gained from the feedback. 
The first, in the steady state $\Mean\sop=\rop^\infty$ reads
\begin{eqnarray}
\begin{array}{lr}
\label{eq:P_d}
\Pow_d=\mathbf{M}\tr\left[\sop\frac{d\Hop_d}{dt}\right]
=\\
\tr\left[\rop^\infty\frac{d}{dt}(-i\epsd\cop^{\dagger}e^{-i\omega_o t}
+\Hc)\right]
=-2\epsd \omega_o c_{\infty}>0,

\end{array}
\end{eqnarray}
where the positivity  indicates power going into (consumed by) the flywheel. 
We restrict our calculations for the deterministic part of feedback, i.e., we replace $\Hop_f(t)$ by its deterministic part 
$\Hop_{f,det}=-i\kapf\ave{\cop}_\sigma\cop\dg+\Hc$.
As was mentioned before, considering the stochastic part 
$\Hop_{f,sto}=-i\kapf/\sqrt{\gamm})\cop\dg d\xi+\Hc$ requires the Stratonovich calculus. 
The power reads
\begin{widetext}
\begin{equation}
\label{eq:p_fdet}
\Pow_{f,det}=\mathbf{M}\tr\left[\sop\frac{d\Hop_{f,det}}{dt}\right]
 =
  \mathbf{M}\tr\left[\sop\frac{d}{dt}(-\kapf\ave{\cop}_\sigma\cop\dg+\Hc)\right]
= 
-i\kapf\mathbf{M}\tr\left[\frac{d\sop}{dt}\cop\right]\ave{\cop\dg}_\sigma+\mathrm{c.c}.
\end{equation}
\end{widetext}

The power in eq.(\ref{eq:p_fdet}) is proportional to the (weighted) mean of the phase drift of the amplitude $\ave{\cop}_\sigma$.
To calculate $d\sop$ we apply the final SME given in  Appendix D. 
The only relevant yield is the unitary rotation $-i\omo\ave{\cop}_\sigma dt$ since the dissipative part does not alter the phase of   $\ave{\cop}_\sigma$
and the  Ito stochastic part will cancel out by the mean operation $\Mean$. Therefore we get
\beq
\label{eq:P_f}
\Pow_{f,det}=-2\kapf \omega_o \mathbf{M}|\ave{\cop}_{\sigma}|^2<0.
\eeq
Negativity means that power is gained (supplied) by feedback. 
Although analytical solutions for similar SMEs such as ours exist \cite{gisin91}, we restrict ourselves to a simple guess.
Using  the Cauchy-Schwartz relation $\mathbf{M}\vert\ave{\cop}_\sigma
\vert^2 \geq \vert\mathbf{M}\ave{\cop}_\sigma\vert^2$, we obtain the lower bound  
$-\Pow_{f,det}\geq 2\kapf\omega_o |c_{\infty}|^2$ 
for the stationary power gained by feedback in the steady-state. 
Hence the overall stationary power satisfies the inequality
\beq 
\label{eq:P_lowerbound}
-\Pow_{det}=-\Pow_d-\Pow_{f,det} \geq 2\omo\epsd^2\frac{-\kapeneg}{(\kapf+\kapeneg)^2}. 
\eeq
The sign is negative and thus the consumable power of the flywheel is positive and bounded from below. 
We conjecture that the contribution of the stochastic part $\Hop_{f,sto}(t)$ of driving can not invalidate the positivity of the consumable power.
\par
We summarize the plausible structure of energy currents (see Fig.\ref{fig:schem}d). 
The steady state energy balance contains five different currents:
\beq
\dot\Ener = \Heat_e + \Heat_m + \Heat_f + \Pow_d + \Pow_f=0.
\eeq 

The heat flowing into the flywheel has two contributions,  the first is from the engine, $\Heat_e$, the second is from the monitoring device,  $\Heat_m$.
Power from the driving field, $\Pow_d$, is also consumed by the flywheel, and serves as an input power activating the flywheel.
This power is overcompensated by the output power $\Pow_f$ realized by the feedback.
In addition, the outflow $\Heat_f$ cools the flywheel, thereby stabilizing it and lowering the entropy produced in the flywheel as a result of the engine and the monitoring operations.
In the case $\beteneg \rightarrow0^-$ of no population inversion in the engine, the heat flow $\Heat_e$ and the consumable power must vanish. 
The work \eqref{eq:Work} stored in the flywheel reaches its minimal, yet positive, value $\Work=\omo \epsd^2/\kapf^2$.
\section{Summary}
Population inversion, corresponding to negative temperature $1/\beteneg$ in a few-level quantum heat engine has been established a long time ago \cite{ramsey56}. 
In this paper we have shown that the heat engine operation is equivalent to a negative temperature heat bath in the standard dynamical sense.
Thus, its influence on the work repository is the typical thermalizing master equation extended to negative temperature $1/\beteneg$. 
\par
Work extraction  is still an outstanding issue because of the spread of thermal and quantum noise over the work repository which in our case is a quantum HO. 
If the HO is replaced by an idealized classical field, all the energy flowing out of the engine can in principle be extracted as power (see Appendix E).
This results from the fact that classical field generates a unitary operation on the system which does not change its entropy.
However, when a work repository is quantized the work exchange is accompanied with heat exchange which may degrade the charging efficiency.
In this paper we introduced a generic novel approach that can be applied to resolve such problems.    
Specifically, we demonstrated the difficulties of storing useful work in a quantum harmonic oscillator. 
Overcoming the unlimited growth of fluctuations, regulating and stabilizing the flywheel is achieved by applying monitoring and feedback control to the system.
\par
The steady-state, the power, and the stored extractable energy of the flywheel are determined analytically. 
While the amount of work stored in the flywheel is  independent of the accuracy of the monitoring, the charging efficiency is optimized for a special ratio between the monitoring and the feedback strength. 
In other words there is a balance between information gained by monitoring the flywheel and the information fed back to the flywheel.
The balance coincides with minimum temperature of the flywheel. 
Breaking this balance implies that the phase space distribution is no longer optimal for work extraction from the flywheel.
\par
This model is a prototype of an analytically tractable model of a quantum heat engine coupled to a single degree-of-freedom work repository, operating continuously in steady state under quantum control. 
Experiments which employ quantum monitoring and feedback strategies are becoming common in recent years \cite{,clerk08,mabuchi04,vijay12,nori14}. 
Future advances in quantum technologies depend on our ability to control and manipulate quantum system. 
A firm theoretical foundation relating systems that are subject to quantum monitoring and feedback control with basic concepts of thermodynamics is still missing. 
\section{Acknowledgments}
We thank Raam Uzdin, Saar Rahav, David Gelbwaser-Klimovsky and Walter Singaram  for fruitful discussions. This work was supported by the Israeli Science Foundation and the Hungarian Scientific Research Fund under Grant No. 103917.
Part of this work was supported by the COST Action MP1209 ``Thermodynamics in the quantum
regime''.  
\par
\section*{Appendix A: Tripartite Heat Engine}
\label{sec:a}
We use an interaction picture  for its convenience especially for our master equations.
The stochastic master equations of monitoring and feedback are presented in the Schr\"{o}dinger picture 
for transparency. Heat flow and power are, as a rule, defined in the Schr\"{o}dinger picture.
We derive the master equation for the harmonic oscillator (HO) subject to the operation of the engine. 
The quantum heat engine is comprised of two two-level systems (TLS's), with the Hamiltonians $\Hop_h = \omh \aop^{\dagger}\aop$ and $\Hop_c= \omc \bop^{\dagger} \bop$.
 The two TLS's are coupled to a hot and a cold heat bath, respectively, at temperatures $T_h>T_c$. 
The dynamics follow the Lindblad-Gorini-Kossakowski-Sudarshan dynamics \cite{lindblad76,gorini276},
in the interaction picture of $\Hop_{h(c)}$ the corresponding master equations read 
\begin{widetext}
\begin{eqnarray}
\label{eq:LGKS}
\begin{array}{ll}
 \frac{d\rop_h}{dt}=
\Gamma_h\left(\aop\rop_h\aop^\dagger-\Herm \aop^\dagger\aop\rop_h\ +e^{-\betah\omh}(\aop^\dagger\rop_h\aop- \Herm  \aop\aop^\dagger\rop_h)\right)\equiv\Lch\rop_h\\
 \frac{d\rop_c}{dt}=
\Gamma_c\left(\bop\rop_c\bop^\dagger-\Herm\bop^\dagger\bop\rop_c\ +e^{-\betc\omc}(\bop^\dagger\rop_c\bop- \Herm\bop \bop^\dagger\rop_c)\right)\equiv\Lcc\rop_c\\
\end{array}
\end{eqnarray}
\end{widetext}
where $\Gamma_{h(c)}$ are the damping rates. 
(In our convention, different from  ref.'s \cite{wisemanbook}, $\Herm$ denotes the Hermitian part of all what stands after it.)
The heat baths bring the TLS's to thermal equilibrium states $\rop_{h(c)}^\infty$
with the occupation numbers $n_{h(c)}=1/(e^{\beta_{h(c)}\omega_{h(c)}}+1)$, and with the inverse temperatures $\beta_{h(c)}=1/T_{h(c)}$, respectively.

The two TLS's are then weekly coupled to a quantum HO of self-Hamiltonian $\Hop_o=\omo\cop^{\dagger}\cop$, via the tripartite Hamiltonian
\begin{equation}
\label{eq:Hint}
 \Kop =-ig\aop\bop^{\dagger}\cop^{\dagger} + H.c.
\end{equation}
We work in resonance, $\omo=\omh-\omc$, and in the weak coupling regime for which a local master equation holds \cite{levy14b}.
The master equation in the interaction picture for the tripartite state $\rop_3$ of the TLS's coupled to the HO is written as 
\begin{equation}
\label{eq:master1}
 \frac{d\rop_3}{dt} = (\mathcal{L} + \mathcal{K})\rop_3,
\end{equation}
with  $\mathcal{L}=\Lch + \Lcc$ and
\begin{equation}
\mathcal{K} \rop_3 = -i[\Kop ,\rop_3]. 
\end{equation}
We will to derive the effective master equation for the HO's state $\ropo$ assuming that the TLS's are initially in their equilibrium states
$\rop_{hc}^\infty=\rop_h^\infty\otimes\rop_c^\infty$ and the initial state of the tripartite system is the product state $\rop_3(0)=\rop_{hc}^\infty\otimes\ropo(0)$.
The solution of the master equation  \eqref{eq:master1} can be written
in the implicit form
\begin{equation}
 \rop_3(t)=\rop_3(0) + \int^{t}_0 ds  e^{\mathcal{L}(t-s)}\mathcal{K} \rop_3(s),
\end{equation}
which we can confirm by taking the time derivative of both sides of the equation, and using the relation $\mathcal{L}\rop_3(0)=0$.
Inserting the above solution into the rhs  of eq.\eqref{eq:master1}, we obtain
\begin{equation}
 \frac{d \rop_3(t)}{dt}=\mathcal{K}\rop_3(0) + (\mathcal{L}+\mathcal{K})\int^{t}_0 ds  e^{\mathcal{L}(t-s)}\mathcal{K} \rop_3(s).
\end{equation}
We assume that $\rop_3(s)\approx \rop_{hc}^\infty\otimes\ropo(s)$.
This assumption is justified when the thermalization time of the TLS's is faster than the time scale in which the system is changed significantly due to coupling (\ref{eq:Hint}). 
Taking the partial trace over the TLS's:
\begin{equation}
 \frac{d \ropo(t)}{dt}=\tr_{hc}[\mathcal{K}\int^{t}_0 ds  e^{\mathcal{L}(t-s)}\mathcal{K} \rop_{hc}^\infty\otimes\ropo(s)].
\end{equation}
Here we have used the relations 
$\tr_{hc}[\mathcal{K}\rop_{hc}^\infty ]=0$ and $\tr_{hc}[\mathcal{L}\int_0^t e^{\mathcal{L}(t-s)} \mathcal{K} \rop_{hc}^\infty]=0$.
Performing the standard Markovian approximations \cite{breuer} we obtain
\begin{equation}
 \frac{d \ropo(t)}{dt}=\tr_{hc}[\mathcal{K}\int^{\infty}_0 ds  e^{\mathcal{L}s}\mathcal{K} \rop_{hc}^\infty\otimes\ropo(t)],
\end{equation}
which can be written explicitly as
\begin{eqnarray}
 \frac{d \ropo}{dt}=-\tr_{hc}\left[\Kop,\int^{\infty}_0 ds  e^{\mathcal{L}s}\left[\Kop , \rop_{hc}^\infty \otimes \ropo\right]\right]
\\ \nonumber
=-\tr_{hc}\int^{\infty}_0 ds \left[\left(e^{\mathcal{L^{\dagger}}s}\Kop \right), \left[ \Kop, \rop_{hc}^\infty\otimes\ropo\right]\right].
\end{eqnarray}
Making use of the relation
\begin{equation}
 e^{\mathcal{L^{\dagger}}s}\Kop  = \Kop\exp[-\half \sum_{l=h,c}\Gamma_l(1+e^{-\beta_l \omega_l})s],
\end{equation}
we have
\begin{widetext} 
\begin{equation}
\frac{d\ropo}{dt}= \frac{(2g)^2}{\sum_{l=h,c} \Gamma_l(1+ e^{-\beta_l \omega_l})}
\left[\langle\aop\aop^\dagger\rangle\!_\infty\langle\bop^\dagger\bop\rangle\!_\infty(\cop\ropo \cop^\dagger-\Herm\cop^\dagger\cop\ropo)
 +\langle\aop^\dagger\aop\rangle\!_\infty\langle\bop\bop^\dagger\rangle\!_\infty(\cop^\dagger\ropo \cop-\Herm\cop\cop^\dagger\ropo) \right]
\end{equation} 
\end{widetext}
where $\ave{\cdot}_\infty$ stands for the expectation value with respect to the TLS's thermal equilibrium states $\rop_{h(c)}^\infty$. 
Finally, the master equation for the HO subject to the engine operation takes the form
\begin{equation}
\label{eq:L_e}
\frac{d\ropo}{dt}\equiv\Lce\ropo=
\Gamma_e(\cop\ropo\cop\dg-\Herm\cop\dg\cop\ropo)
      +\Gamma_e e^{-\beteneg\omo}(\cop\dg\ropo\cop-\Herm\cop\cop\dg\ropo),
\end{equation} 
where
\begin{equation}
\Gamma_e=(2g)^2\frac{(1-n_h)^2(1-n_c)n_c}{\Gamma_h(1-n_c)+\Gamma_c(1-n_h)},
\end{equation} 
and the output temperature of the heat-engine
\begin{equation}
\beteneg=\frac{\betah\omh - \betc\omc}{\omh-\omc}
\end{equation} 
which is a function of the TLS's excitation energies and temperatures only.
We operate the system as a heat engine, i.e.  $T_h/T_c > \omh/\omc >1$, 
the effective temperature is negative, i.e. $1/\beteneg < 0$, and the HO will not reach a stable asymptotic state, as we show below. 
The master equation \eqref{eq:L_e} together with the Hamiltonian $\Hop_o$ yield closed evolution equations for the mean
amplitude $\ave{\cop}_t$ as well as for the occupation $\ave{\cop^\dagger\cop}_t$:
\begin{eqnarray}
\label{eq:c_t}
\frac{d\ave{\cop}_t}{dt}&=&-\left(\kapeneg+i\omo\right)\ave{\cop}_t\\
\label{eq:n_t}
\frac{d\ave{\cop^\dagger\cop}_t}{dt}&=&-2\kapeneg\ave{\cop^\dagger\cop}_t+\Gamma_e e^{-\beteneg\omo}
\end{eqnarray} 
where
\beq
\kapeneg = \half\Gamma_e\left(1-e^{-\beteneg\omo}\right)<0
\eeq
is the standard amplitude damping constant. This time it is negative
since $\beteneg<0$ therefore both $\ave{\cop}_t$ and $\ave{\cop^\dagger\cop}_t$ diverge exponentially with time.
In particular, a thermal state remains thermal, the temperature is increasing
exponentially as can be shown by the simple solution of \eqref{eq:n_t} for
the occupation. Note, however, that our model is only valid in the weak coupling regime where the thermalization time is shorter than the internal time scale. This implies that the occupation must be limited by
\begin{equation}
\label{eq:weak}
g\sqrt{\ave{\cop^\dagger\cop}+1}\ll\Gamma_{h(c)} \left(1+e^{-\beta_{h(c)}\omega_{h(c)}}\right).
\end{equation} 

\section*{Appendix B: External Driving}
Coupling the HO to a resonant oscillating external field. Via such \emph{driving} one would expect to extract power.
Consider the time-dependent Hamiltonian in the Schr\"{o}dinger picture,
\begin{equation}
\label{eq:Hext}
\Hop_d(t)=-i\epsd\left(\cop^{\dagger}e^{-i\omo t} -\cop e^{i\omo t}\right)
\end{equation}
where $\epsd>0$.
In the interaction picture, the master equation \eqref{eq:L_e} is modified by an additional static Hamiltonian:
\begin{equation}
\label{eq:L_e_ext}
\frac{d\ropo}{dt}=\Lce\ropo-\epsd[\cop^{\dagger} -\cop,\ropo].
\end{equation} 
Now the rhs of eq. \eqref{eq:c_t} of the mean amplitude acquires an additional term $-\epsd e^{-i\omo t}$.
This allows for an exceptional  stationary solution of constant amplitude with the rotating phase:
\begin{equation}
\label{eq:c-const}
\ave{\cop}_t=-\frac{\epsd}{\kapeneg}e^{-i\omo t}=\mathrm{const}\times e^{-i\omo t}.
\end{equation} 
This solution is unstable since all neighbouring solutions exponentially diverge with $t$. 
As to the occupation $\ave{\cop^\dagger\cop}_t$,
 the rhs of eq.\eqref{eq:n_t} acquires the additional linear term
$-\epsd(\ave{\cop^\dagger}_t-\ave{\cop}_t)$ hence the occupation remains exponentially divergent,
there is no steady state solution under external driving. 
The stability issue of the HO is still not resolved. 
\section*{Appendix C: Monitoring}
Continuous measurement, i.e.: \emph{monitoring}, is  the first task towards feedback control on the system \cite{wisemanbook}.  
Here we consider the time-continuous measurement of both quadratures 
$\hat{x}=\frac{1}{\sqrt{2}}(\cop^\dagger +\cop)$ and $\hat{y}=\frac{i}{\sqrt{2}}(\cop^\dagger -\cop)$ of the HO.
Generalizing the result of \cite{diosi88} for monitoring simultaneously $\hat{x}$ and $\hat{y}$, 
we can write the following stochastic master equation (SME) in the Schr\"{o}dinger picture for the density matrix $\sop$ conditioned on both measurement signals $\bar{x},\bar{y}$:
\begin{widetext}
\begin{eqnarray}
\label{eq:SME}
d\sop&=&-i\left[\Hop_o,\sop\right]dt-\frac{\gamm}{8}\left[\hat{x},\left[\hat{x},\sop\right]\right]dt-\frac{\gamm}{8}\left[\hat{y},\left[\hat{y},\sop\right]\right]dt
+ \Herm\gsqrt\left(\hat{x}-\left\langle\hat{x}\right\rangle_\sigma\right)\sop d\xi_{x}
     + \Herm\gsqrt\left(\hat{y}-\left\langle\hat{y}\right\rangle_\sigma\right)\sop d\xi_{y}.
\end{eqnarray}
\end{widetext}
All expectation values $\ave{\cdot}_\sigma$ are understood in the stochastic conditional state $\sop$. 
The measurement signals satisfy 
\begin{equation}
\label{eq:signals_xy}
\bar{x}dt=\ave{\hat{x}}_\sigma dt+\frac{d\xi_x}{\gsqrt},~~~~
\bar{y}dt=\ave{\hat{y}}_\sigma dt+\frac{d\xi_y}{\gsqrt}.
\end{equation}
Here $d\xi_x,d\xi_y$ are Ito increments of independent standard Wiener processes, satisfying 
\beq
(d\xi_x)^2 =(d\xi_y)^2 = dt,~~d\xi_x d\xi_y=0,~~
\Mean d\xi_x=\Mean d\xi_y=0,
\eeq
with the symbol $\Mean$ for stochastic mean, and $\gamm$ for the measurement  strength. 
(Note that we changed $\gamm$ in ref. \cite{diosi88} for $\gamm/2)$.  
We can return to complex notation, i.e.,  we rewrite the above equations in terms of $\cop,\cop^\dagger$ and the corresponding complex signal $\cbar=(\bar{x}+i\bar{y})/\sqrt{2}$.
We define the complex Wiener increment 
\beq
d\xi=\frac{d\xi_x+i d\xi_y}{\sqrt{2}}
\eeq
that satisfies
\beq
(d\xi)^2=(d\xi^\ast)^2=0,~~~~~d\xi^\ast d\xi=dt,~~~~~\Mean d\xi=\Mean d\xi^\ast=0.
\eeq
The SME \eqref{eq:SME} of the conditional state becomes
\begin{widetext}
\begin{eqnarray}
d\sop&=&-i\left[\Hop_o,\sop\right]dt
+\frac{\gamm}{4}\left(\cop\ropo\cop^\dagger-\Herm\cop^\dagger\cop\sop
                                            +\cop^\dagger\ropo\cop-\Herm\cop\cop^\dagger\sop\right)
+ \gsqrt\Herm\left(\left(\cop-\ave{\cop}_\sigma\right)d\xi^{*}+\Hc\right)\sop\nonumber\\
&\equiv&-i\left[\Hop_o,\sop\right]dt +\Lcm\sop dt
+ \gsqrt\Herm\left(\left(\cop-\ave{\cop}_\sigma\right)d\xi^{*}+\Hc\right)\sop
\end{eqnarray}
\end{widetext}
Eqs.\eqref{eq:signals_xy} of the real signals take the following form for the complex signal:
\begin{equation}
\label{eq:signal_c}
\bar{c}dt=\ave{\cop}_\sigma dt+\frac{d\xi}{\gsqrt}.
\end{equation}
Applying this time-continuous measurement to the HO which is coupled to the heat engine
and driven by the external field,  cf. eq. \eqref{eq:L_e_ext}, we get the following SME:
\begin{widetext}
\beq
\label{eq:smemonitor}
d\sop=-i[\Hop_o,\sop]dt+(\Lce+\Lcm)\sop dt-\epsd[\cop^\dagger e^{i\omo t}-\cop,\sop]dt
+\gsqrt \Herm[(\cop-\ave{\cop}_\sigma)d\xi^\ast+\Hc]\sop.
\eeq 
\end{widetext}
The state $\sop$ of the HO is the conditioned state on the measured signal \eqref{eq:signal_c},
its stochastic mean is the unconditional density matrix: $\Mean\sop=\ropo$.
Taking the stochastic mean $\Mean$ of both sides of the SME, 
we are left with the master equation of the unconditional state:
\begin{equation}
\frac{d\ropo}{dt}= (\Lce+\Lcm)\ropo-\epsd[\cop^\dagger-\cop,\ropo].
\end{equation}
As a result of the measurement, additional heat flows into the oscillator, the damping
rate becomes  $\Gamma_e+\gamm$ and the inverse "temperature" $\beteneg$ is
modifies but remains negative.  
The exceptional steady amplitude \eqref{eq:c-const} exists with the
modified parameters, but it is unstable like all other solutions.   
\section*{Appendix D: Feedback Control}
Using the measured signal \eqref{eq:signal_c}, we control the state of the HO in the vicinity of the constant rotating
amplitude set by the external driving, in such a way that we get a true stable
steady state. Consider the following \emph{feedback} Hamiltonian in the Schr\"{o}dinger picture:
\beq
\Hop_f(t)=-i\kapf\cbar(t)\cop\dg +\Hc.
\eeq
Here $\kapf$ is the feedback strength.
We apply the feedback \cite{diosi94} on top of the monitored evolution described by eq.(\ref{eq:smemonitor}):
\begin{equation}
\label{eq:drho}
\sop+d\sop\rightarrow e^{-i\Hop_f dt}(\sop+d\sop) e^{i\Hop_f dt}.
\end{equation}
Expanding the rhs into a series, keeping first order terms in $dt $ and keeping in mind that  $\vert d\xi\vert^2=dt$, 
the terms that are left for evaluation are: $-i\left[H_fdt,\sop\right]$, $-i\left[H_fdt,g\sop\right]$ and $-\half \left[H_fdt,\left[H_fdt,\sop\right]\right]$. 
The final SME including feedback reads:
\begin{widetext}
\begin{equation}
\label{eq:smecontrol}
d\sop  =  -i[\Hop_o,\sop]dt+(\Lce+\Lcm+\Lcf)\sop dt-\epsd\left[\cop^\dagger e^{i\omo t}-\cop e^{-i\omo t},\sop\right]dt
  -\frac{\kapf}{\gsqrt}\left[\cop^\dagger d\xi-\cop d\xi^{*},\sop\right]
         +\gsqrt\Herm\left[\cop-\ave{\cop}_\sigma)d\xi^{*}+\Hc\right]\sop.
\end{equation}
\end{widetext}
The dissipative contribution of the feedback reads:
\begin{eqnarray}
\label{eq:L_f}
\Lcf\sop=
 \left(\frac{\kapf^2}{\gamm}+\kapf\right)\left(\cop\sop\cop\dg-\Herm\cop\dg\cop\sop\right)
\\ \nonumber
+\left(\frac{\kapf^2}{\gamm}-\kapf\right)\left(\cop\dg\sop\cop-\Herm\cop\cop\dg\sop\right).
\end{eqnarray}
For $\kapf>\gamm$ this corresponds to a thermal bath of positive temperature.  Entering the regime $0<\kapf<\gamm$,
the cooling effect of $\Lcf$ within the sum  $\Lce+\Lcm+\Lcf$ becomes enhanced although 
$\Lcf$ ceases to be a mathematically correct dissipator in itself.
Taking the stochastic mean over eq. (\ref{eq:smecontrol}) we obtain the master equation of the unconditional state 
which in the interaction picture takes this form : 
\begin{equation}
\label{eq:me}
\frac{d\ropo}{dt}=(\Lce+\Lcm+\Lcf)\ropo-\epsd\left[\hat{c}^{\dagger}-\hat{c},\ropo\right].
\end{equation}

What we have got for the HO dynamics is the following:
The HO is excited by the negative temperature ($1/\beteneg$) bath $\Lce$ due to population inversion, heated by the infinite temperature bath  $\Lcm$ due
to noise of monitoring, and cooled by the feedback $\Lcf$.
On top of this, the external driving shifts the Hamiltonian $\Hop_o$.  
We write the full master equation \eqref{eq:me} in a compact form:
\begin{eqnarray}
\begin{array}{lr}
\label{eq:me_final}
\frac{d\ropo}{dt}=
 \Gamma\left(\cop\ropo\cop\dg-\Herm\cop\dg\cop\ropo\right)
+\Gamma e^{-\beta\omo}\left(\cop\dg\ropo\cop-\Herm\cop\cop\dg\ropo\right)
\\
~~~ - ~\epsd[\cop\dg-\cop,\ropo]
\end{array}
\end{eqnarray}
where $\Gamma$ and $\beta$ are determined by
\beq
\label{eq:rate_d}
\Gamma=\Gamma_e+\frac{\gamm}{4}+\frac{\kapf^2}{\gamm}+\kapf,
\eeq
\beq
\label{eq:rate_u}
e^{-\beta\omo}\Gamma=\Gamma_e e^{-\beteneg\omo}+\frac{\gamm}{4}+\frac{\kapf^2}{\gamm}-\kapf.
\eeq
\par 
We turn the effective temperature $\beta$ positive by choosing the feedback strength above the following threshold:  
\beq
\kapf > -\kapeneg=\half\Gamma_e\left(e^{-\beteneg\omo}-1\right).
\eeq
Note that the driving on the rhs of the master equation \eqref{eq:me_final} can be absorbed into the standard
thermal dissipator at (inverse) temperature $\beta$ if we displace $\cop,\cop\dg$ by a suitable real number.
Accordingly, the master equation  \eqref{eq:me_final} must have a unique stationary state which is the following displaced 
thermal state of the HO:
\beq
\label{eq:final_state}
\ropo^{\infty} =\mathcal{N}\exp\left[-\beta\omo\left(\cop-c_\infty\right)^{\dagger}\left(\cop-c_\infty\right)\right]
\eeq   
with the static real displacement in interaction picture:
\begin{equation}
\label{eq:c-const-displacement}
c_\infty=-\frac{\epsd}{\kapf+\kapeneg}<0.
\end{equation} 
In the Schr\"{o}dinger picture  the stationary state is a thermal state with the rotating displacement:
\beq
\label{eq:final_state-Sch}
\ropo^{\infty} \Rightarrow \mathcal{N}\exp\left[-\beta\omo\left(\cop-c_\infty e^{-i\omo t}\right)^{\dagger}\left(\cop-c_\infty e^{-i\omo t}\right)\right].
\eeq   
Hence the mean amplitude rotates, its phase is shifted by $-\pi/2$ with respect to the external driving:
\beq
\label{eq:mean_ampl}
\ave{\cop}_\infty=c_\infty e^{-i\omo t}.
\eeq   
The average population is the Planckian thermal value plus the yield of displacement:
\beq
\ave{\cop^{\dagger}\cop}_\infty =  \frac{1}{e^{\beta\omo}-1} +\vert c_\infty\vert^2 \equiv  n_o +\vert c_\infty\vert^2 ,
\eeq
we use the redundant expression   $\vert c_\infty\vert^2$ for $c_\infty^2$ to capture an occasionally 
different phase convention of driving. Both terms on the rhs diverge at the edge of the regime of operation
$\kapf+\kapeneg\rightarrow+0$ where the model breaks down because it violates the weak coupling condition \eqref{eq:weak}.   
\section*{Appendix E: Power Extraction Via Classical periodic Field }
\label{sec:e}
Steady state power extraction \emph{without} storing work is possible by just by driving the engine directly without the flywheel.
Power is gained by  amplification of a classical rotating field in resonance with the two TLS's. 
The interaction Hamiltonian is given by  $\Kop(t)=-i\epsilon \left(\aop \bop^{\dagger}e^{i\omo t} -\aop^{\dagger} \bop e^{-i\omo t}\right)$.
For weak driving the master equation for the two TLS's, $\rop_{hc}$, in the interaction picture of $\Hop_h$ and $\Hop_c$ is,
\begin{equation}
\label{eq:mse_classical_field }
\frac{d\rop_{hc}}{dt}=-\left[\epsilon\left(\aop\bop^{\dagger}-\aop^{\dagger}\bop \right),\rop_{hc}  \right]+\mathcal{L}_h\rop_{hc}+\mathcal{L}_c\rop_{hc},
\end{equation}
where $\mathcal{L}_{h(c)}$ are defined in (\ref{eq:LGKS}). 
The master equation (\ref{eq:mse_classical_field }) posses a unique stationary state.  
The stationary output power,
\begin{widetext}
\begin{equation}
-\mathcal{P}^{\infty}=\frac{4\epsilon^2 \omo (n_h-n_c)}{4\epsilon^2\left( \Gamma_h^{-1}(1-n_h)+\Gamma_c^{-1}(1-n_c)\right)+\Gamma_h\left( 1+e^{-\beta_h\omh}\right)+\Gamma_c\left( 1+e^{-\beta_c\omc}\right)  }>0,
\end{equation} 
\end{widetext}
is positive. 
This implies that steady state power extraction can be obtained from a periodically driven field.
Note that for strong driving there is also a steady state power extraction from the engine. 
Nevertheless, the master equation (\ref{eq:mse_classical_field }) must be modified. 
Derivation of a master equation  driven by a strong periodic field for such a model can be found in \cite{levy212}.

\end{document}